\documentclass[referee,onecolumn] {aa}
\usepackage {txfonts}
\usepackage {graphicx}

\begin{document}

\title{Influences of magnetic coupling process on the spectrum of
a disk covered by the corona}

\author{R.-Y Ma \and D.-X Wang \and X.-Q Zuo}
\offprints{D. X. Wang, \email{dxwang@hust.edu.cn}}
 \institute{Department of Physics, Huazhong University of Science and Technology,Wuhan 430074,
 China}

\date{Received     /  Accepted }

\abstract {Recently, much attention has been paid to the magnetic
coupling (MC) process, which is supported by very high emissivity
indexes observed in Seyfert 1 galaxy MCG-6-30-15 and GBHC XTE
J1650-500. But the rotational energy transferred from a black hole
is simply assumed to be radiated away from the surrounding
accretion disk in black-body spectrum, which is obviously not
consistent with the observed hard power-law X-ray spectra.} {We
intend to introduce corona into the MC model to make it more
compatible with the observations.}{We describe the model and the
procedure of a simplified Monte Carlo simulation, compare the
output spectra in the cases with and without the MC effects, and
discuss the influences of three parameters involved in the MC
process on the output spectra.}{It is shown that the MC process
augments radiation fluxes in the UV or X-ray band. The emergent
spectrum is affected by the BH spin and magnetic field strength at
the BH horizon, while it is almost unaffected by the radial
profile of the magnetic field at the disk.}{Introducing corona
into the MC model will improve the fitting of the output spectra
from AGNs and GBHCs.}


 \maketitle



\section{INTRODUCTION}

Tenuous hot plasma (corona) is commonly believed to exist in the
inner region of the accretion flow, which is produced and heated
by magnetic activities, such as reconnection and flare etc. (e.g.
\cite{Liang77}; \cite{Galeevetal79}). The influences of the disk
corona on the emergent spectrum from active galactic nuclei (AGNs)
and galactic black-hole candidates (GBHCs) have been discussed by
many authors (e.g. Haardt {\&} Maraschi 1991, hereafter
\cite{Haardt91}; \cite{Haardt93}; \cite{Field93}; \cite{Esin99};
\cite{Zhangetal00}; \cite{Liuetal02}). Corona explains the
power-law X-ray spectra very well, and reprocessing of the coronal
X-rays by cold disk gives rise to the observed emission lines
naturally, in which iron $K\alpha $ fluorescence line provides us
a diagnostic of the geometry of the accretion flow and the
property of the spacetime around the black hole (BH). However,
only gravitational energy of the accreting matter is invoked to
heat the disk and its corona in previous works.

Recently, much attention has been paid to the magnetic coupling
(MC) process in which energy is transferred from a rotating BH to
its surrounding disk (\cite{Blandford99}; \cite{Putten99};
\cite{Li00}; Li \cite{Li02a}; Wang et al. \cite{W02}; Wang et al.
2003a, b, hereafter \cite{W03a} and \cite{W03b}, respectively),
which is in fact a variant of the Blandford-Znajek (BZ) process.
The existence of the MC process has been supported by very high
emissivity indexes observed in Seyfert 1 galaxy MCG-6-30-15 and
the GBHC XTE J1650-500, which cannot be explained by the standard
accretion disk (Wilms et al. \cite{Wilmsetal01};
\cite{Milleretal02}; \cite{Li02b}).

However, in the previous MC models all the energies that deposit
on the disk, both the gravitational energy of the accreting matter
and the rotational energy transferred from the BH, are simply
assumed to be radiated away in black-body spectrum, which is
obviously not consistent with the observed hard power-law X-ray
spectra. In this paper we intend to introduce corona into the MC
model and discuss the spectrum affected by the MC process. It is
shown that the MC process augments the radiation fluxes, while
corona makes MC model more reasonable and more consistent with the
observations than the previous MC model.

The paper is arranged as follows: Sect.2 gives the description of
our model. Sect.3 shows the procedure and results of our
simplified Monte Carlo simulation and in Sect.4 we discuss the
applications of our model to observations. Throughout this paper
the geometric units $G = c = 1$ are used.

\section{DESCRIPTION OF OUR MODEL}

\begin{figure}
\begin{center}
\includegraphics[width=7cm]{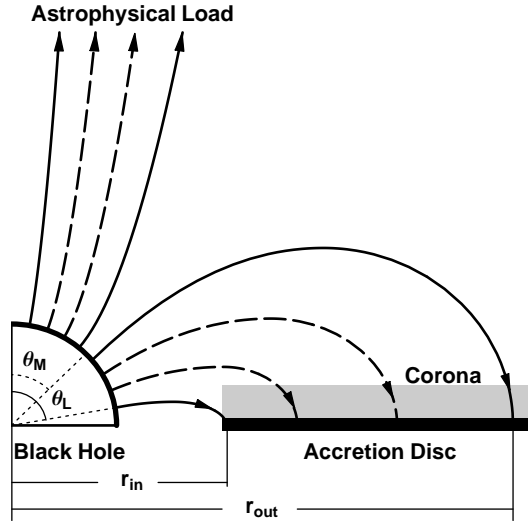}
\end{center}
\caption{Poloidal profile of the large-scale magnetic field and
geometry of corona} \label{fig1}
\end{figure}

The poloidal profile of the large-scale magnetic field and the
geometry of the corona are illustrated in Fig.~\ref{fig1}. The
central BH is fast rotating and surrounded by axisymmetric
magnetosphere. It is assumed that the disk is geometrically thin
and optically thick, while the corona is geometrically thick and
optically thin. The poloidal magnetic field is assumed to be
constant on the horizon and it varies as a power-law with the
radial coordinate on the disk, i.e.,

\begin{equation}
\label{eq1}
B_H = const,
\quad
B_D \propto r^{ - n}.
\end{equation}

In Fig.1 the angle $\theta _M $ indicates the angular boundary
between the open and closed field lines at the BH horizon, which
can be determined by the mapping relation derived in \cite{W03b}.
The angle $\theta _L $ is the lower angular boundary of the closed
field lines, and it is supposed to be less than $0.5\pi $ to avoid
the singularity of the closed field lines at the equatorial plane.
The magnetic field powers the jet via the BZ process and transfers
energy between the BH and the disk via the MC process.

Both large- and small-scale magnetic fields are involved in our
model though the latter is not shown in Fig.~\ref{fig1}. The
large-scale magnetic field plays a key role in transferring energy
and angular momentum from a fast-rotating BH to its surrounding
disk or remote astrophysical load in the MC and BZ processes,
respectively. While the small-scale magnetic field energizes the
corona above the disk by virtue of some physical processes, such
as magnetic reconnection and buoyancy (\cite{Haardt91}). Moreover,
the large-scale field can be produced from the small-scale field
created by dynamo processes (e.g. \cite{Tout96}). Livio et al.
(\cite{Livioetal99}) argued that the strength of the large-scale
field threading a black hole is very weak if the field is created
in the thin disk, because the ratio of the large-scale field to
the small-scale field is $\sim H / R \ll 1$, where $H$ is the disk
thickness and $R$ is the radial size of the field. This greatly
confines the BZ and MC powers. Since the thickness of the corona
is about the radius of the disk, the large-scale magnetic fields
created by dynamo processes in the corona are significantly
stronger than those in the thin disk (\cite{Cao04}), and the
strong BZ and MC powers are maintained. For simplicity, we suppose
that these two kinds of magnetic fields exist independently.

The energy deposited in the corona and disk consists of two parts:
(i) the gravitational energy of the accreting matter, and (ii) the
rotational energy of the BH transferred in the MC process. Some
authors (e.g. \cite{Haardt91}; \cite{Merloni02}) have mentioned
physical processes in which the gravitational energy converts to
the thermal energies of the corona and disk. However, nobody has
proposed any mechanism to describe how the rotational energy of
the BH transferred in the MC process dissipates to date. In this
paper, we deal with this issue based on the following two
considerations.

(i) Based on BH magnetosphere theory (\cite{MT82}; \cite{TPM}; Li
\cite{Li02a}; \cite{W03a}), the disk can be regarded as an
electromotive force (EMF) in an equivalent circuit due to its
rotation, which is opposite to the BH EMF. If the angular velocity
of the BH is greater than that of the disk, Poynting flux would
flow from the BH to the disk (\cite{W03b}). Although the disk is
commonly supposed to be a perfect conductor, its interior
resistance may not be zero. Part of the MC power may heats the
disk, just as Poynting flux gets into the resistance and heats it
in a circuit;

(ii) The flux of the angular momentum transferred magnetically
from the BH to the disk decreases rapidly with disk radius (see
Eq.(\ref{eq4}) below), and it results in the differential rotation
of the accreted matter. Thus the electromagnetic energy converts
to thermal dissipation, just as the gravitational energy of
accreting matter converts to thermal dissipation in disk
accretion.

Detailed discussions of these physical processes are beyond the
scope of this paper, and we assume that both the gravitational
energy of the disk and the rotational energy of the BH can
dissipate into thermal energy of the corona and disk.

The scenario of the emergent spectrum is described as follows.
Thermal soft seed photons are emitted from the disk, and when they
cross the rarefied hot corona, some may go through the corona
directly without being scattered, some may escape after one or
several times of scattering, and others may be down-scattered to
the disk by the corona. The fate of the X-ray photon that returns
to the disk depends critically upon its energy $E$: (a) if $E >
100\mathrm{keV}$, the photon loses its energy by Compton recoil,
(b) if $E < a\mbox{ }few  \mbox{ keV}$, the photon is absorbed by
ionization or free-free absorption, and (c) it may be both Compton
scattered and bound-free absorbed for  $\mbox{ } a \mbox{ }
few\mathrm{ keV} < E < 100\mathrm{keV}$ (\cite{Field93}).
Bound-free absorption or ionization results in a vacancy in the
inner shell of an atom, which leads to a fluorescence photon or
the ejection of an Auger electron (for details see George \&
Fabian 1991, hereafter \cite{George91}). Unscattered photons form
the component of the UV/soft X-ray band, photons escaping the
corona after inverse Compton scatterings constitute the power-law
X-ray spectrum, while reprocessed photons by the disk produce
fluorescence lines and a reflection hump.

Here we neglect cyclo-synchrotron radiation since it is less
important than Compton radiation in the slab corona
(\cite{Matteoetal97}; \cite{Ghisellinietal98}). We also neglect
the bremsstrahlung radiation, assuming that the corona is tenuous
enough.

From the above scenario it can be found that part of the energy
dissipated in the corona is conveyed to the disk via hard X-rays,
i.e., the disk is heated by three kinds of energies: (1)
gravitational energy of the accreting matter, (2) rotational
energy of the BH transferred to the disk and (3) the absorbed
energy of X-rays coming from the corona. Since the energy of the
X-rays comes eventually from the gravitational and rotational
energies, we suppose the fraction of the total energy dissipate in
the disk to be $\eta$, which includes the contribution of the
downward X-rays. As shown in \cite{Haardt91}, the soft emission is
derived almost only from absorption and reprocessing of the
high-energy flux impinging on the disk. Roughly half of the
Comptonized photons in the corona irradiate and heat the disk, and
the rest half radiate away. So we take $\eta=0.5$ in calculations.

Additionally, for simplicity of simulation, the disk is assumed to
be ``cold'', with hydrogen and helium being fully ionized, but all
other elements being neutral or weakly ionized. Moreover, the
energetic and spatial distributions of free electrons in the
corona is postulated to be thermal and homogeneous.

\section{SIMULATION OF DISK SPECTRUM}

\subsection{ Multicolor Black-Body Spectrum of the Disk Affected by the MC Process}

Based on the conservation laws of mass, angular momentum and
energy, the local radiation flux from the disk around a rotating
BH is given by Page \& Thorne (\cite{Page74}) and Li
(\cite{Li02a}),

\begin{equation}
\label{eq2} F  = \frac{\dot{M}_D}{4\pi r}f + ( - \frac{d\Omega _D
}{rdr})(E^ + - \Omega _D L^ + )^{ - 2}\int_{r_{ms} }^r {(E^ + -
\Omega _D L^ + )H rdr},
\end{equation}

\noindent where $\dot{M}_D$, $\Omega _D$, $r_{ms}$, $E^+$ and
$L^+$ are respectively the accretion rate, the angular velocity of
the disk, the radius of the marginally stable orbit, the specific
energy and angular momentum at the disk radius $r$. The function
$f$ is related to the radiation flux due to disk accretion, and
$H$ is the flux of angular momentum transferred between the BH and
the disk (\cite{W03a}),

\begin{equation}
\label{eq4}
 H(a_\ast;\xi,n)/H_0=\left\{
 \begin{array}{c@{\quad , \quad}c}
   A(a_\ast,\xi)\xi^{-n} & 1<\xi<\xi_{out} \\ 0 & \xi>\xi_{out}
 \end{array} \right.,
\end{equation}

\begin{equation}
\label{eq5}
 A(a_ * ,\xi) = \frac{a_ * (1 - \beta)}{2\pi \left[ {2\csc ^2\theta -( 1 - q)}
\right]}\sqrt {\frac{1 + a_
* ^2 \chi _{ms}^{ - 4} \xi ^{ - 2} + 2a_ * ^2 \chi _{ms}^{ - 6} \xi ^{ -
3}}{1 - 2\chi _{ms}^{ - 2} \xi ^{ - 1} + a_ * ^2 \chi _{ms}^{ - 4}
\xi ^{ - 2}}} ,
\end{equation}

\noindent where $H_0 \equiv (B^p_H)^2 M$, $\beta$ is the ratio of
the angular velocity of the disk to that of the BH, i.e. $\beta
\equiv \Omega _D / \Omega _H $, $\xi = r / r_{ms} $ is a radial
parameter of the disk defined in terms of $r_{ms} $ , and $\chi
_{ms} \equiv \sqrt {r_{ms} / M} $.


The relation between the magnetic field and the accretion rate is
given by Moderski et al. (\cite{Moderskietal97}) based on the
balance between the pressure of the magnetic field on the horizon
and the ram pressure of the innermost parts of an accretion flow,
i.e.,

\begin{equation}
\label{eq3}
 (B^p_H)^2 / (8\pi) = P_{ram} \sim \rho c^2\sim \dot {M}_D /(4\pi
 r_H^2),
\end{equation}

\noindent where $r_H$ is the radius of the horizon. From
Eq.(\ref{eq3}) we can define $F_0 \equiv
(B^p_H)^2=2\dot{M}_D/[M^2\left(1+\sqrt{1-a_*^2} \right)^2]$.

For the presence of corona the dissipated power in the unit area
of the disk, $F_d$, is related to $F(r)$ by

\begin{equation}
\label{eq6}
F_d (r) = \eta F(r),
\end{equation}

\noindent where $F(r)$ is the local radiation flux expressed by
Eq.(\ref{eq2}). According to Stefan-Boltzmann law we have the
local effective temperature on the disk expressed by

\begin{equation}
\label{eq7} T_d (r) = (F_d / \sigma_\mathrm{SB} )^{1 / 4},
\end{equation}

\noindent where $\sigma_\mathrm{SB}$ is the Stefan-Boltzmann
constant. The local radiation spectrum is defined by Planck
function:

\begin{equation}
\label{eq8} B_v (r) = \frac{2 h }{c^2}\frac{v^3}{\exp \left[h v/
k_B T_d (r)\right] - 1}.
\end{equation}

\noindent Thus the multicolor black-body spectrum of the disk is

\begin{equation}
\label{eq9}
L_v = \int_{r_{in} }^{r_{out} } {B_v (r)2\pi rdr} .
\end{equation}

\subsection{Monte Carlo Simulation}

If the disk is covered by corona, we need to resolve the radiative
transfer in the hot corona to get the emergent spectrum. Radiative
transfer has been computed in different ways. One approach is to
solve the radiative transfer equation either numerically or
analytically (e.g. \cite{Sunyaev80}; \cite{Poutanen96}). Another
approach is the Monte Carlo simulation (e.g. Pozdnyakov et al.
1983, hereafter \cite{Pozdnyakovetal83}; \cite{Gorecki84};
\cite{George91}; \cite{Sternetal95}; Hua 1997, hereafter
\cite{Hua97}; Yao et al. \cite{Yaoetal05}). The emergent spectrum
can be also obtained by some approximate semi-analytical formulae
(e.g. \cite{Zdziarski86}; \cite{Hua95}). In this paper we
calculate the spectrum by using the Monte Carlo simulation.

Compared with H97, photon absorption in the disk is taken into
account in our model, which makes the simulation slightly more
complex. The steps of the simulation are: (i) sample a seed photon
including its position, energy and direction; (ii) draw a value
for its free path and test whether it can leave the disk or
corona; (iii) simulate the interaction of the photon with the
medium; (iv) repeat steps (ii), (iii) till the photon leaves the
system of the corona and disk.

\subsubsection{Sampling the seed photons}

The probability density of the seed photon can be written as

\begin{equation}
\label{eq10} p(r,v) = 2\pi rdr \cdot F_v (r) / L,
\end{equation}

\noindent where $F_v(r)$ is the flux density at $r$, $L \equiv
\int_{r_{in} }^{r_{out} } {L(r)dr} $ is the total luminosity of
the disk, and $L(r)dr = 2\pi rF_d (r)dr$ is the luminosity of the
ring $r\sim r + dr$. Obviously, $p(r,v)$ can be expressed as
follows,

\begin{equation}
\label{eq11}
 p(r,v) = \frac{L(r)dr}{L} \cdot \frac{2\pi r \cdot
F_v (r)dr}{L(r)dr} = \frac{L(r)dr}{L} \cdot \frac{F_v (r)}{F_d
(r)} =  \tilde{p}(r)B_v(r)
\end{equation}

\noindent where $\tilde{p}(r)$ is the probability of a photon
emitted in the ring $r\sim r + dr$, which can be sampled by
tabulation. The Planck function $B_v(r)$ can be sampled in the way
described in \cite{Pozdnyakovetal83}.

\subsubsection{Drawing the free path}
For the photon scattered in the corona, the probability it travels
at least an optical depth $\tau $ is $e^{ - \tau }$, so the
optical depth that the photon travels between the i-th and
(i+1)-th scatterings can be drawn with $\tau _i = - \ln \lambda $,
where $0 \le \lambda \le 1$ is a random number corresponding to a
random event. Then the free path of the photon can be drawn with $
\frac{\tau _i }{n_e \sigma } = \frac{ - \ln \lambda }{\sigma }
\cdot \frac{\sigma _\mathrm{T} H_c }{\tau _c }$, where $n_e $,
$\sigma $ and $\sigma_\mathrm{T}= 6.65\times 10^{ -
24}\mathrm{cm}^{2}$ are the number density of the electrons,
cross-sections of scattering and Thomson Scattering, respectively.
The parameters $H_c $ and $\tau _c $ are the vertical height and
optical depth of the corona, respectively. Because the corona is
very hot, electrons in it are relativistic, which means that
$\sigma $ depends not only on the energy of the photon but also on
the energy and direction of the electron. So the cross-section
averaged over the distribution of the electrons in \cite{Hua97} is
used to draw the free path of the photon.

Since the disk is assumed to be ``cold'', with hydrogen and helium
being fully ionized, and all other elements being neutral or
weakly ionized, hard X-ray photons irradiating the disk from the
corona can be absorbed by the atoms, as well as being scattered by
the free electrons. In this case, there are two viable ways to
draw the free path of the photon: (i) choose the lesser of the two
free paths that are drawn with $\sigma _a$ and $\sigma _s$, where
$\sigma _a$ and $\sigma _s$ are respectively the cross sections of
absorption and scattering(\cite{George91}); (ii) draw the free
path with $\sqrt{\sigma_a(\sigma_a+\sigma_s)}$, and determine the
interaction by drawing a random number $\lambda$ and comparing it
with the probability that a free path ends with absorption, i.e.,
$\zeta=\frac{\sigma_a}{\sigma_a+\sigma_s}$. If $\lambda \leq \zeta
$, the photon will be absorbed and otherwise be scattered
(\cite{Rybicki79}). We follow the first way in our simulations.
Since electrons in the disk is nonrelativistic, $\sigma _{s} $ can
be given by the Klein-Nishina formula:

\begin{equation}
\sigma _{s} = \frac{3\sigma _\mathrm{T} }{4} \cdot
\frac{1}{x}\left[ {\left( {1 - \frac{4}{x} - \frac{8}{x^2}}
\right)\ln (1 + x) + \frac{1}{2} + \frac{8}{x} - \frac{1}{2(1 +
x)^2}} \right],
\end{equation}

\noindent where $x = 2hv / m_e c^2$ is the energy of the incident
photon in unit of electron-rest-energy. The cross section of
absorption $\sigma _{a}$ is taken from Morrison {\&} McCammon
(\cite{Morrison83}).

As the free path is known, the position that the photon arrives
before the next interaction can be calculated. If the photon is
outside the corona-disk system, it will escape away from the
system and its energy and direction are recorded. But if the
photon transfers from the disk to the corona or inverse, the point
where the trajectory of the photon cross the interface between the
disk and corona should be regarded as the next initial position of
the photon in calculations (see \cite{Hua97} for details).

\subsubsection{Simulating the interaction}

If the photon is scattered in the corona or disk, we can first
sample an electron from thermal distribution, and then calculate
the energy and direction of the scattered photon, following the
procedure described in \cite{Pozdnyakovetal83}.

The bound-free absorption of hard X-rays by the atoms in the disk
will lead to ionization and vacancy, and induce emission of
fluorescence lines with the probability called fluorescence yield,
Y, rather than ejection of Auger electrons. In the simulations, if
the photon is absorbed by a certain atom or ion, we can draw a
random number $\lambda $ ($0 \le \lambda \le 1$) and compare it
with the corresponding fluorescence yield. If $\lambda < Y$, an
emission line is brought out, whose direction can be sampled from
the isotropic distribution. If $\lambda > Y$, an Auger electron is
sent out, the photon vanishes and its trajectory ends.

Because of the large abundance and cross-section of absorption of
iron, only the 6.4keV Fe K$\alpha$ fluorescence line is considered
in our simulations. If the energy of the absorbed photon is less
than the iron K-shell absorption edge, i.e. $E < 7.1\mathrm{keV}$,
no emission line is produced, while Fe K$\alpha$ lines emanate
with $Y = 0.34$ for $E > 7.1 \mathrm{keV}$.

\subsection{ Parameters of Corona and Results of Simulations}

Optically thin thermal Compton scattering spectra are well
represented by power-laws with exponential cut-offs, and the
spectral index can be approximately given by (e.g.
\cite{Rybicki79})

\begin{equation}
\label{eq13} \alpha = - \frac{\ln P}{\ln A_1}.
\end{equation}

\noindent In Eq.(\ref{eq13}) $A_1 = 1 + 4\Theta + 16\Theta ^2$ is
the average photon energy amplification per scattering, where
$\Theta = kT_c / m_e c^2$ with $T_c $ being the coronal
temperature. The average scattering probability $P$ for a uniform
slab is given by \cite{Zdziarskietal94} as

\begin{equation}
\label{eq14} P = 1 + \frac{\exp ( - \tau )}{2}\left(
{\frac{1}{\tau } - 1} \right) - \frac{1}{2\tau } + \frac{\tau
}{2}E_1 (\tau ),
\end{equation}

\noindent where $E_1 $ is the exponential integral
(\cite{Pressetal92}).

Observations of Seyfert galaxies show that their spectral indexes
and the coronal Thomson optical depths are about 0.9 and 1,
respectively (\cite{Haardt91}; \cite{Zdziarski98}), and we take
$\alpha = 0.9$, $\tau = 1$  in simulations. From Eq.(\ref{eq13}),
we estimate the temperature of the corona to be $T_c \approx
60\mathrm{keV}$.

In the following calculations, the radii of the inner and outer
edges of the corona and disk are taken as $ r_{ms}$ and $20r_{ms}
$, respectively. The half heights of the disk and the corona are
assumed to be $H_d $=0.02$r_{ms} $ and $H_c $=$r_{ms} $,
respectively. The value of $H_d \ll r_{ms}$ does not affect the
final results. The radial range of the disk is taken to be rather
small, because the MC power concentrates in the innermost region
of the disk, and it can be easily verified that at least 80{\%} of
the radiation is emanated within $20r_{ms} $.

Since the typical magnetic field of the supermassive and stellar
BHs are respectively about $10^4$ G and $10^8$ G, we take the
middle value, $10^6$ G, in simulations. A large number ($10^9)$ of
photons are traced to get one spectrum in simulations, and the
result of $10^9$ events costs about 100 minutes in running the
program on a PC with 2.8 GHz Pentium 4 CPU. All the spectra in
this paper are given in forms of curves rather than histogram to
make the curves look smooth.

The simulation results are given in Fig.~2. The spectrum of the
disk-corona system comprises three components, the black-body
spectrum formed by unscattered photons, the power-law spectrum
formed by photons that escape from the corona after several times
of inverse Compton scatterings, and the reflected spectrum
characterized by iron fluorescence line and reflection hump.

\begin{figure}
\centerline{\includegraphics[width=7cm]{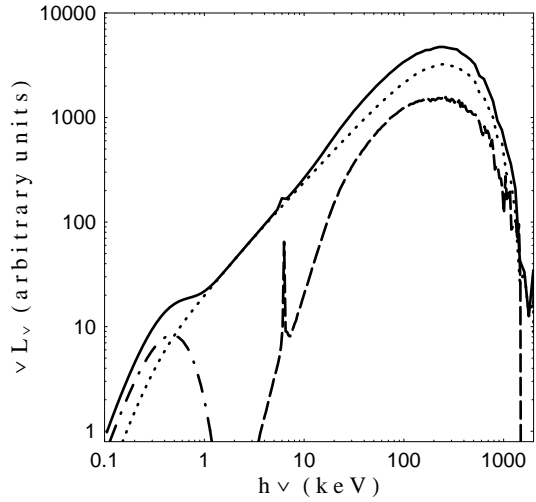}}
 \label{fig2}
\caption{The spectrum of the disk-corona system (solid line) and
its components. The black-body, power-law and reflected components
are shown in dot-dashed, dotted and dashed lines, respectively.
The parameters $a_\ast = 0.998,\mbox{ }n = 3$  and $B =
10^6\mathrm{G}$ are taken.}
\end{figure}

In order to study the effects of the MC process and corona, the
spectra of the disk-corona system with and without the MC process
and those with and without corona are simulated as shown in
Fig.~3.

\begin{figure}
\centerline{\includegraphics[width=7cm]{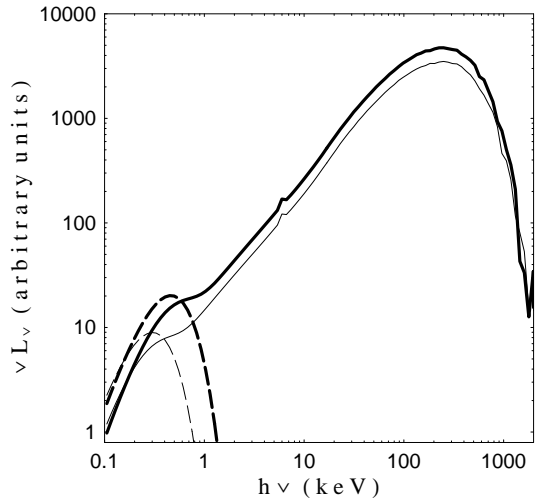}} \label{fig3}
\caption{Multicolor spectra of the disk and the simulated spectra
of the disk-corona system. Thick and thin lines correspond to the
spectra with and without the MC process, respectively. Solid and
dashed lines correspond to the spectra with and without corona,
respectively.The parameters $a_\ast = 0.998,\mbox{ }n = 3$  and $B
= 10^6\mathrm{G}$ are taken.}
\end{figure}


\begin{figure}
\centerline{\includegraphics[width=6.5cm]{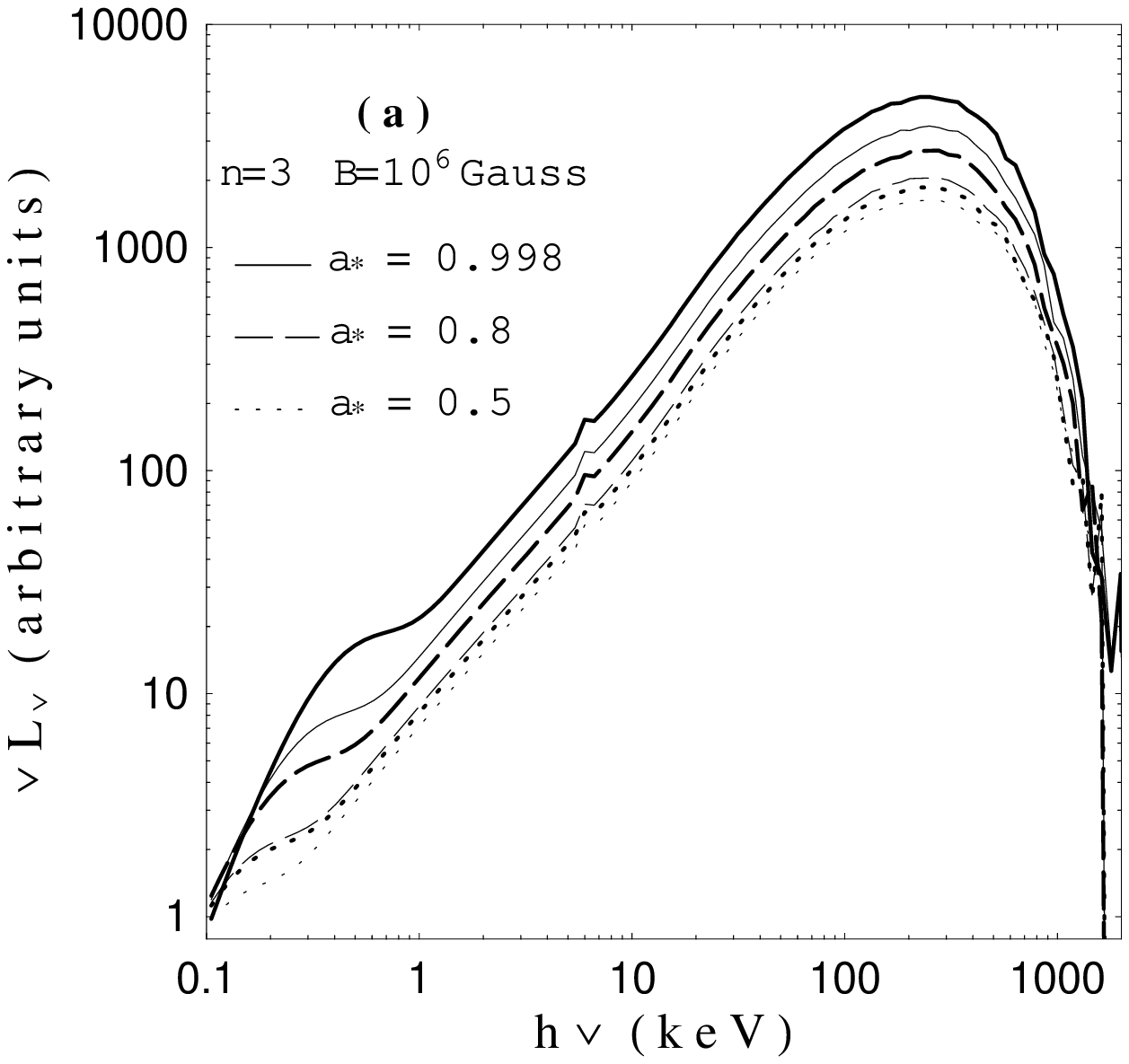}}
\centerline{\includegraphics[width=6.5cm]{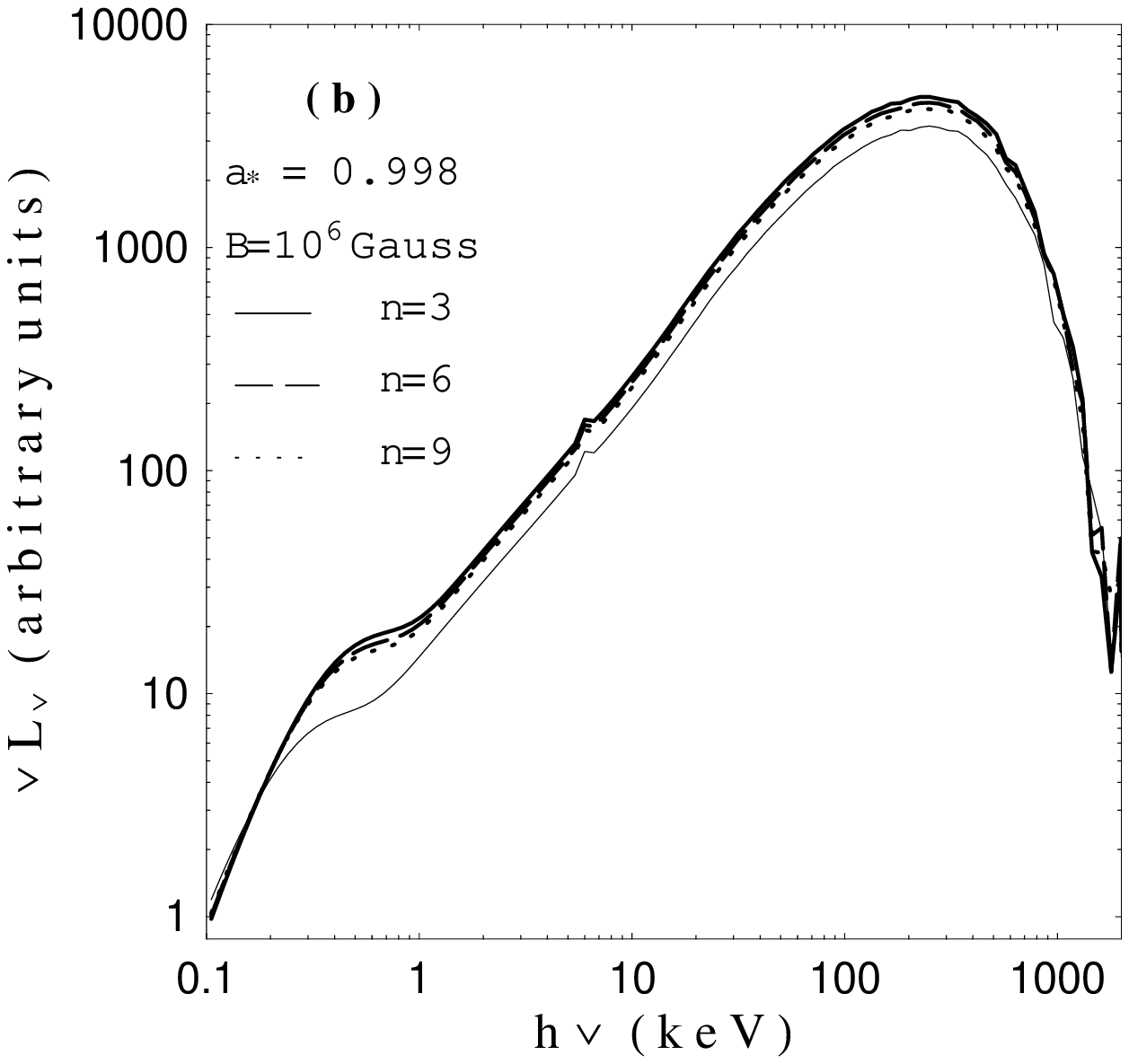}}
\centerline{\includegraphics[width=6.5cm]{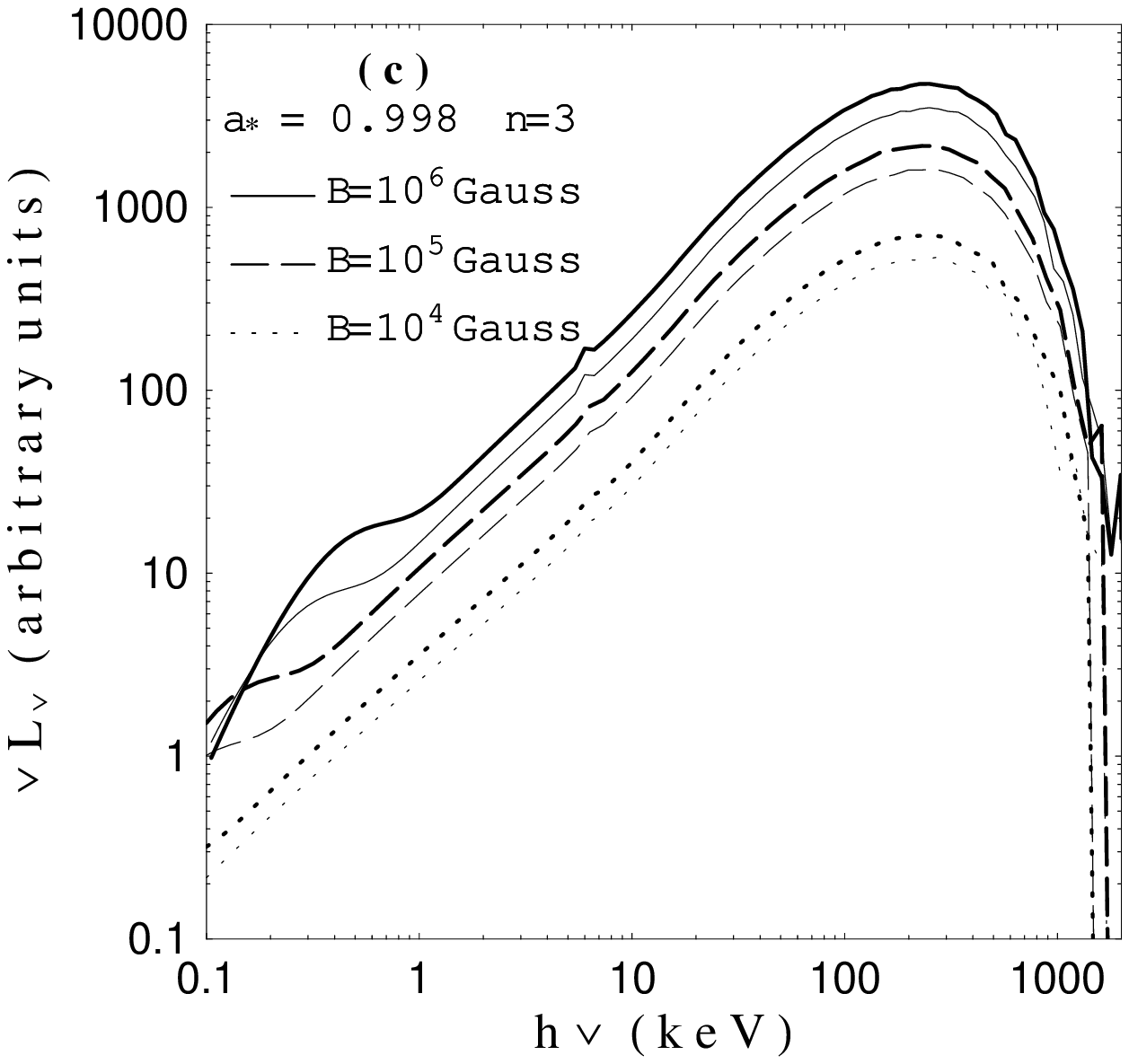}}

\label{fig4}
 \caption{Spectra corresponding to different values of
the parameters $a_ * $, $n$ and $B$ in (a), (b) and (c),
respectively. Thick and thin lines correspond to the states with
and without the MC process, respectively.}
\end{figure}

From Fig.~3 we find: (i) the disk can at most emit soft X-rays
with $E < a few\mathrm{ keV}$ without corona, while the spectrum
becomes much more complex and much harder when corona exists; (ii)
the MC process can increase the flux and extend the spectrum to
higher energy when the corona does not exist, but it can only
augment the flux and has no influence on the cutoff energy and the
index of the spectrum.

The fact that the cutoff energy and index of the spectrum are
indifferent with the MC process when corona exists is because the
spectral cutoff $hv_{cut} $ and index $\alpha $ are determined by
$\Theta $ and $\tau $, which are independent of the MC process in
our model.

Spectra with different $a_\ast $, $n$ and $B$ are given in Fig.~4,
from which we find (i) the fluxes increase with $a_\ast $ and $B$
no matter the MC process exists or not; (ii) the parameter $n$ has
little influence on the flux.

It can be easily explained why the fluxes increase with $a_\ast $
and $B$ no matter the MC process exists or not. For pure accretion
without the MC process, as $a_\ast $ increases the inner edge of
disk approaches the BH horizon, and more gravitational energy can
be released if the accretion rate is the same. So the fluxes of
the disk-corona system increase with $a_*$. According to
Eq.(\ref{eq3}), stronger magnetic field corresponds to a greater
accretion rate $\dot {M}_D $, and thus higher fluxes. At the
presence of the MC process, higher fluxes are produced for greater
$a_\ast $ and $B$ since the MC power increases with both $a_\ast $
and $B$. The flux is almost unaffected by $n$ because the MC power
is not significantly affected by $n$.

\section{ APPLICATION TO ASTROPHYSICS}

In this paper corona is introduced into our model, and it is
helpful to interpret the following observations.

\subsection{Jet}

The same as our previous model, the BZ and MC processes can
coexist in the present model, provided that the parameters $a_\ast
$ and $n$ are greater than some critical values, and their powers
are derived by using a modified equivalent circuit and expressed
as follows (\cite{W03b}),

\begin{equation}
\label{eq15}
 P_{BZ}/P_0 = 2a_ * ^2 \int_0^{\theta _M }
{\frac{k(1 - k)\sin ^3\theta d\theta }{2 - (1 - q)\sin ^2\theta }}
,\mbox{ }0 < \theta < \theta _M,
\end{equation}

\begin{equation}
\label{eq16}
 P_{MC} /P_0 = 2a_ * ^2 \int_{\theta _M }^{\theta _L }
{\frac{\beta (1 - \beta)\sin ^3\theta d\theta }{2 - (1 - q)\sin
^2\theta }} ,\mbox{ }\theta _M < \theta < \theta _L,
\end{equation}

\noindent where $P_0 \equiv \left( {B_H^p } \right)^2M^2$.

An important feature of our model is the strength of the BZ power
relative to the MC power can be adjusted by virtue of the angle
$\theta _M $, which is the angular boundary between the open and
closed field lines at the BH horizon as shown in Fig.1. As argued
in \cite{W03b}, $\theta _M $ is eventually determined by the
parameters $a_\ast$ and $n$ contained in the mapping relation
between the angular coordinate on the BH horizon and the radial
coordinate on the disk.

Combining accretion with the MC process, the luminosity of the
disk
 can be expressed as

\begin{equation}
\label{eq17}
 L = (1 - E_{ms} )\dot {M}_D + P_{MC}.
\end{equation}

The first term in Eq.(\ref{eq17}) comes from disk accretion, and
the second term from the contribution of the MC process. The
quantity $E_{ms}$ is the specific energy of the accreting matter
corresponding to the marginally stable orbit and it reads
(\cite{Novikov73})

\begin{equation}
\label{eq18} E_{ms} = \frac{\left( {1 - 2\chi _{ms}^{ - 2} + a_ *
\chi _{ms}^{ - 3} } \right)}{\left( {1 - 3\chi _{ms}^{ - 2} + 2a_
* \chi _{ms}^{ - 3} } \right)^{1 / 2}} .
\end{equation}

Our model can be applied to the observations of some sources,
fitting the jet power in radio band driven by the BZ process and
fitting the bolometric luminosity driven by disk accretion and the
MC process.

Here we apply our model to fit the observational data of 3C 273.
The BH mass contained is estimated as $6.59\times 10^9M_ \odot $
by reverberation method, the bolometric luminosity $L$ is
estimated to be $\sim 10^{47}\mathrm{erg} \cdot \mathrm{s}^{ - 1}$
(\cite{Paltani05}), and the minimum jet power is given as
$P_{jet}^{\min } = 6.61\times 10^{45}\mathrm{erg} \cdot
\mathrm{s}^{ - 1}$ (Celotti et al. \cite{Celottietal97}). If the
magnetic field at the BH horizon is considered to be $B_H^p \simeq
10^4M_9^{ - 1 / 2} \mathrm{G}$ (Beskin \cite{Beskin97}), we can
estimate some parameters of our model as $a_\ast \simeq 0.76$, $n
\simeq 6.13$ and $\theta _M \simeq 48^ \circ $ by combining Eqs.
(\ref{eq15})--(\ref{eq18}) with the relation (\ref{eq3}).

The spectral index, $\alpha \approx 0.77$, and luminosity of the
X-rays in the range of 3-10keV, $L_{3 - 10} \approx 0.81\times
10^{46}\mathrm{erg} \cdot \mathrm{s}^{ - 1}$ can be derived from
Page et al. (\cite{Pageetal04}), where some X-ray observational
results of 3C273 are listed. Combining Eq.(\ref{eq13}) with the
assumption that $\tau = 1$, we have $T_c \simeq 73\mathrm{keV}$.
It can then be obtained by Monte Carlo method that $L_{3 - 10}
\approx 0.72 \times 10^{46}\mathrm{erg} \cdot \mathrm{s}^{ - 1}$,
which is a little less than the observational data. Since 3C273 is
a well known radio-loud quasar with a jet showing superluminal
motion, the X-rays emitted by synchrotron self-Comptonization of
high-energy electrons in the jet should be considered. Moreover,
the degree of ionization of the disk is supposed to be low in our
model, which means higher absorption in the range $\sim10$keV. If
these reasons are considered, the luminosity of the X-rays could
be well fitted with our model.

\begin{figure}
\centerline{\includegraphics[width=8cm]{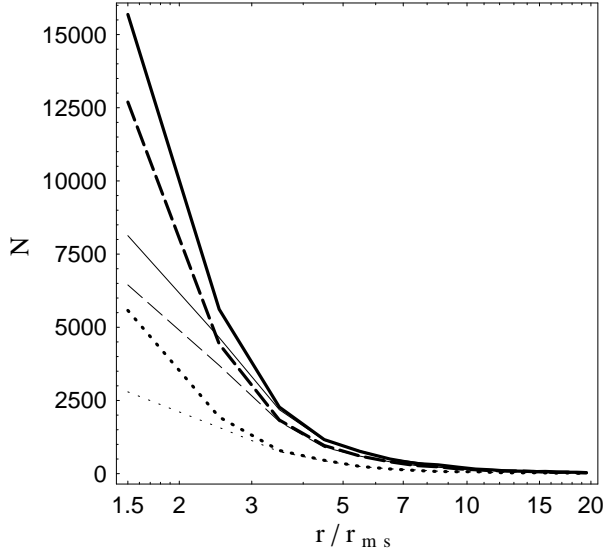}} \label{fig5}
\caption{The numbers of iron fluorescence photons at different
radii and different incline angles. Solid lines, dashed lines and
dotted lines correspond to $\mu = \cos i = 0.95, 0.75, 0.35$,
respectively. Thick lines and thin lines correspond to the cases
with and without the MC effects, respectively. The parameters,
$a_\ast = 0.998, n = 3$ and $B = 10^6\mathrm{G}$ are given.}
\end{figure}

\subsection{ Steep Emissivity Index}
Wilms et al. (\cite{Wilmsetal01}) assumed that the reflected flux
has an emissivity profile $\propto r^{ - \beta }$ in fitting the
2-10keV observational data of Seyfert I galaxy MCG-6-30-15, and
found a very interesting result. The emissivity profile is very
steep with $\beta = 4.3 - 5.0$, which is too high to be explained
by the standard accretion disk model.

In \cite{W03a} and \cite{W03b}, we explained the very steep
emissivity profile by the local radiation flux emitted in
black-body spectrum by virtue of the MC process. However, the
spectra of previous models cannot extend to $\sim$ 10 keV for
AGNs, which is not consistent with the observations. This
inconsistency can be removed in the MC model with corona.

Taking Fe K$\alpha$ fluorescence as an example, we find in Fig.~5
that the photon numbers produced in our model reduce more quickly
with the disk radius than those produced in the standard disk with
corona. It turns out that the very steep emissivity produced in
the MC process could not be changed at the presence of the corona.

\subsection{QPOs}

Wang et al.(\cite{W03c}) explained kilohertz quasi-periodic
oscillations (QPOs) in BH binaries by virtue of rotating hotspots
arising from non-axisymmetric magnetic fields. When corona is
introduced into our model, although the rotational energy of the
BH is shared by the disk and the corona, i.e., the temperature of
the hotspot on the disk decreases, the total flux from the region
of higher magnetic field remains unchanged. So, as the
high-magnetic-field region rotates, the fittings to the frequency
of the QPOs are still valid. An interesting phenomenon different
with previous model is that we can not see a hotter spot in the
corona because of the cooling of more soft photons from the disk.
Additionally, the introduction of corona into our model is helpful
to interpret QPOs which are usually observed in steep power-law
state (\cite{McClintock03}).

Although the introduction of the corona into our model is helpful
to fitting the observations, our model needs to be improved in
some aspects. For example, the more reasonable geometry and
parameters of the corona should be adopted (e.g.
\cite{Haardtetal94}; \cite{Liuetal02}), the cooling of synchrotron
radiation should be considered, and gravitational effects on the
trajectories of photons need to be taken into account since it
cannot be ignored in the inner region of the disk around a rapidly
spinning BH. We shall deal with these problems in future works.

\begin{acknowledgements}
This work is supported by the National Natural Science Foundation
of China under Grant Numbers 10373006, 10573006 and 10121503. The
anonymous referee is thanked for his (her) helpful comments and
suggestions on the improvement of our manuscript. RYM would like
to thank Dr. B. F. Liu and Dr. Y. S. Yao for helpful discussions.
\end{acknowledgements}

\end{document}